\documentclass[epj]{webofc}
\usepackage[varg]{txfonts}   
\usepackage{multirow}
\usepackage[english]{babel}
\usepackage{amsmath}
\usepackage{amssymb}
\usepackage{epsfig}
\usepackage{graphics,psfrag,rotating}
\usepackage{graphicx}
\usepackage{dcolumn}
\usepackage{bm}
\usepackage{slashbox}
\usepackage{array,multirow}
\usepackage{subfigure}

\usepackage{epsfig}
\usepackage{graphicx}
\usepackage{dcolumn}
\usepackage{bm}
\usepackage{color}

\newcommand{\pythia}{{\ttfamily PYTHIA}\xspace}
\newcommand{\herwig}{{\ttfamily HERWIG}\xspace}

\newcommand{\Ep}{E_{\bar p}}

\woctitle{RICAP-14 The Roma International Conference on Astroparticle Physics}
\begin{document}
\title{Highlights on gamma rays, neutrinos and antiprotons \\ from TeV Dark Matter.}
%
%

\author{Viviana Gammaldi \inst{1}\fnsep\thanks{\email{vivigamm@ucm.es}} 
}

\institute{University Complutense Madrid
          }

\abstract{%
        It has been shown that the gamma-ray flux observed by HESS from the J1745-290 Galactic Center source is well fitted as the secondary gamma-rays photons generated from Dark Matter annihilating into Standard Model particles in combination with a simple power law background.
        The neutrino flux expected from such Dark Matter source has been also analyzed.
          The main results of such analyses for 
         $50$ TeV Dark Matter annihilating into $W^+W^-$ gauge boson and preliminary results for antiprotons are presented. 
         
}
\maketitle
\section{Introduction}
Multimessenger astroparticle study is fundamental for Dark Matter (DM) indirect search. Signatures of DM annihilating or decay in astrophysical sources may be observed in cosmic-ray fluxes by Cerenkov telescopes such as VERITAS, HESS, MAGIC and CTA; neutrino telescopes such as ANTARES or IceCube; satellites such as PAMELA, AMS and Fermi or ballon experiments like CAPRICE or BESS \cite{indirect}.  
The secondary products of annihilation and decay of DM particles
contribute to the cosmic-rays differential flux at the Earth as

\begin{equation}
\frac{d \Phi_{\text{cr-DM}}}{dE}=\eta_{\text{cr}}\cdot\sum^2_{a=1} \sum^{\text{ SM channels} }_i \frac{\zeta^{(a)}_i}{a}
\frac{dN^{(\text{cr})}_{i}}{dE}
 \cdot \frac{\kappa^{(a)}_{\text{cr}}}{4 \pi M^a},
 \label{phigen}
\end{equation}

where:

\begin{itemize}
\item $\eta_\text{cr}$ depends on the secondary particles  of interest (observed cosmic-ray flux) and its propagation.
\item the total flux is given by decay  ($a=1$) or annihilation ($a=2$) events of DM particles into the $i$-th Standard Model (SM) particle (annihilation/decay channel). The $\zeta$ factor discerns between these two cases: $\zeta^{(1)}_i=1/\tau_i^{\text{decay}}$ and $\zeta^{(2)}_i=\langle\sigma_i v\rangle$ are respectively the inverse of the decay time and and thermal averaged annihilation cross section times velocity. The probability that DM annihilates or decays into the $i$-th channel depends on the nature of DM.
 \item the differential number $dN_i^{(\text{cr})}/dE$ of cosmic-rays produced at the source by subsequent events of annihilation or decay of SM particles is simulated by means of Monte Carlo events generator software, such as \pythia or \herwig. Some uncertainties may be introduced in the evaluation of both the $\zeta^{(a)}_i$ and $\kappa_\text{cr}$ factor due to the choice of the Fortran or C++ versions of \pythia or \herwig software, as discussed in \cite{MC} for gamma-rays.
\item the $\kappa_\text{cr}$ factor depends on the astrophysics of DM distribution as well as on the cosmic-rays propagation.
For \textit{neutral} cosmic-rays (n-cr) it is the astrophysical factor 

\begin{equation}
\kappa^{(a)}_{\text{n-cr}}\equiv{\langle J \rangle}_{\Delta\Omega}= \frac{1}{\Delta\Omega}\int_{\Delta\Omega}\text{d}\Omega\int_0^{l_{max}(\Psi)} \rho^{(a)} [r(l)] dl(\Psi)\,.
\end{equation}

Here, $\rho(r)$ is the DM density in the halo  and $l$ is the distance from the Sun to any point in the Galaxy disk and halo. The radial distance $r$ of any point in the halo is measured from the GC,
and is related to $l$ by $r^2 = l^2 + D_\odot^2 -2D_\odot l \cos \Psi$, where $D_\odot \simeq 8.5$ kpc is the distance from the Sun
to the center of the Galaxy. The distance from the Sun to the edge of the halo in the direction $\theta$ from the GC is
$l_{max} = D_\odot \cos \theta + \sqrt{r^2-D_\odot^2 \sin \theta}$. For \textit{neutral} particles,  \textit{directional} observations are achievable. In this case, the flux must be averaged over the solid angle of the detector, that is typically of order of $\Delta \Omega = 2 \pi ( 1 - \cos \Psi )$, being $\Psi$ the angular resolution of the telescope.\\
For \textit{charged} cosmic-rays (c-cr), \textit{directional} observations are \textit{not} feasible. In fact, charged particles observed in a given direction might have been originated everywhere in the sky. In this case the $\kappa_\text{c-cr}$ factor is proportional to a diffusion term
\begin{equation}
\kappa_\text{c-cr}^{(a)}\equiv \left(\frac{\rho_\odot}{M}\right)^{(a)}R_{\text{c-cr}}(r_\odot, E).
\end{equation}

The diffusion factor at the position of the sun $R_{\text{c-cr}}(r_\odot, E)$ for \textit{charged} cosmic-ray ($e^\pm, p^\pm$ etc.. ) is the solution of a diffusion equation that depends on the particle of interest and DM distribution. It describes the diffusion of particles in the Galaxy and the production of secondary cosmic-rays due to the interaction with the Interstellar Medium (ISM). The final flux at the position of the Earth also includes Solar magnetic field effect \cite{antipro, Perko}.

\end{itemize}

 In Section \ref{sec-2}, I present the main results of the cosmic-rays analysis of the prospective DM source at the GC: the gamma-rays cut-off detected by HESS at the Galactic Center and the prospective neutrino flux are reviewed. New constraints obtained from the antiprotons study are also presented in Section \ref{sec-2}, while the conclusions are in Section \ref{sec-3}. 

\section{Cosmic rays from TeV Dark Matter at the Galactic Center}
\label{sec-2}
In previous works it has been shown that the collection of data from HESS telescope from the J1749-290 source at the Galactic Center (GC) is well fitted by model independent DM annihilating into SM channels \cite{HESS}. Such study provides gamma-rays fits with DM masses between $1-100$ TeV. While annihilation in lepton channel is excluded at the $99\%$ confidence level, gamma-rays secondary flux produced by DM annihilating into quark and bosons channels well reproduce the data. Here, Eq.(\ref{phigen}) for the cosmic-rays differential flux at the Earth is discussed for heavy DM particle of $\approx 50$ TeV annihilating into $W^+W^-$ channel. Extra-dimensional particles as branons may justify TeV DM annihilating in boson channels \cite{branons, branonsgamma}.  \\      

\begin{figure}[bt]
\begin{center}
\epsfxsize=5cm
\resizebox{\columnwidth}{!}
{\includegraphics{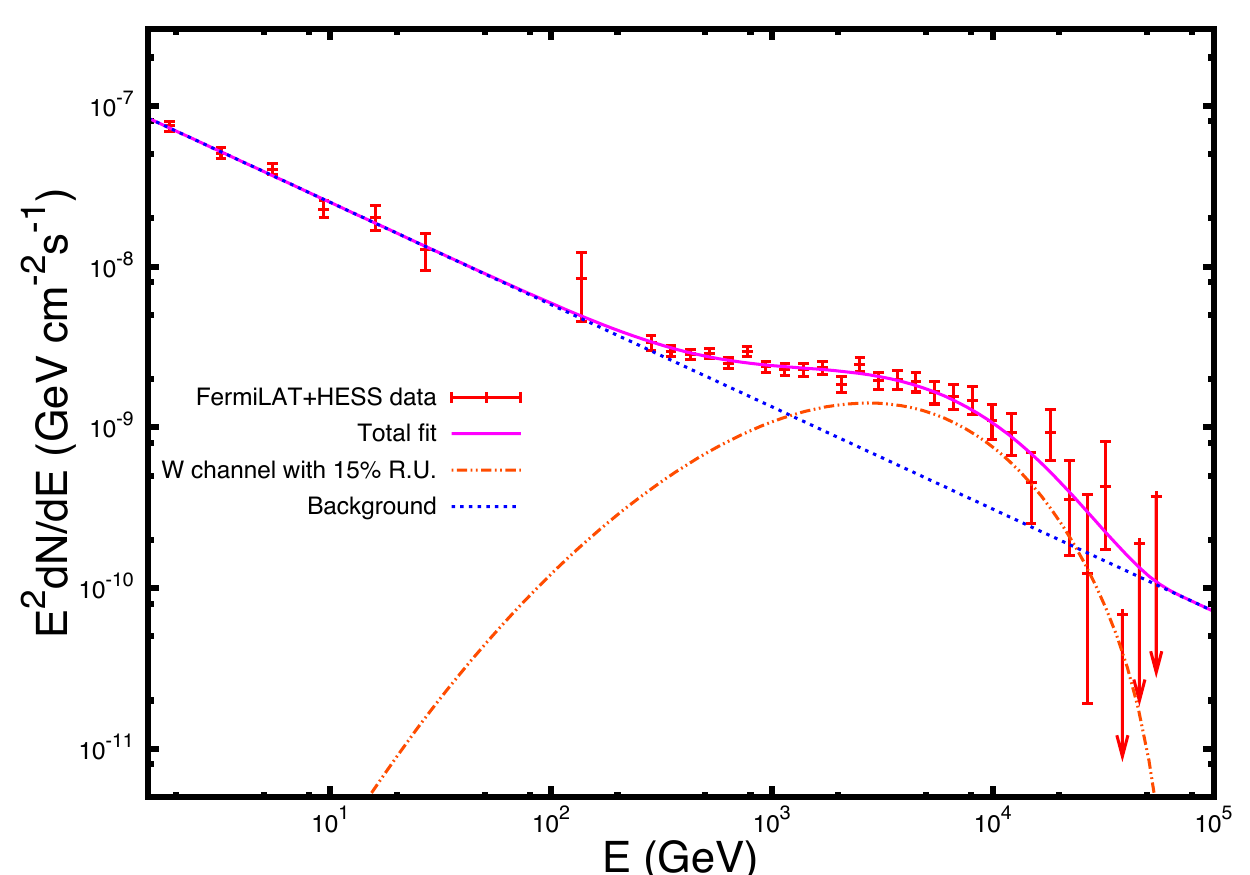}{\includegraphics{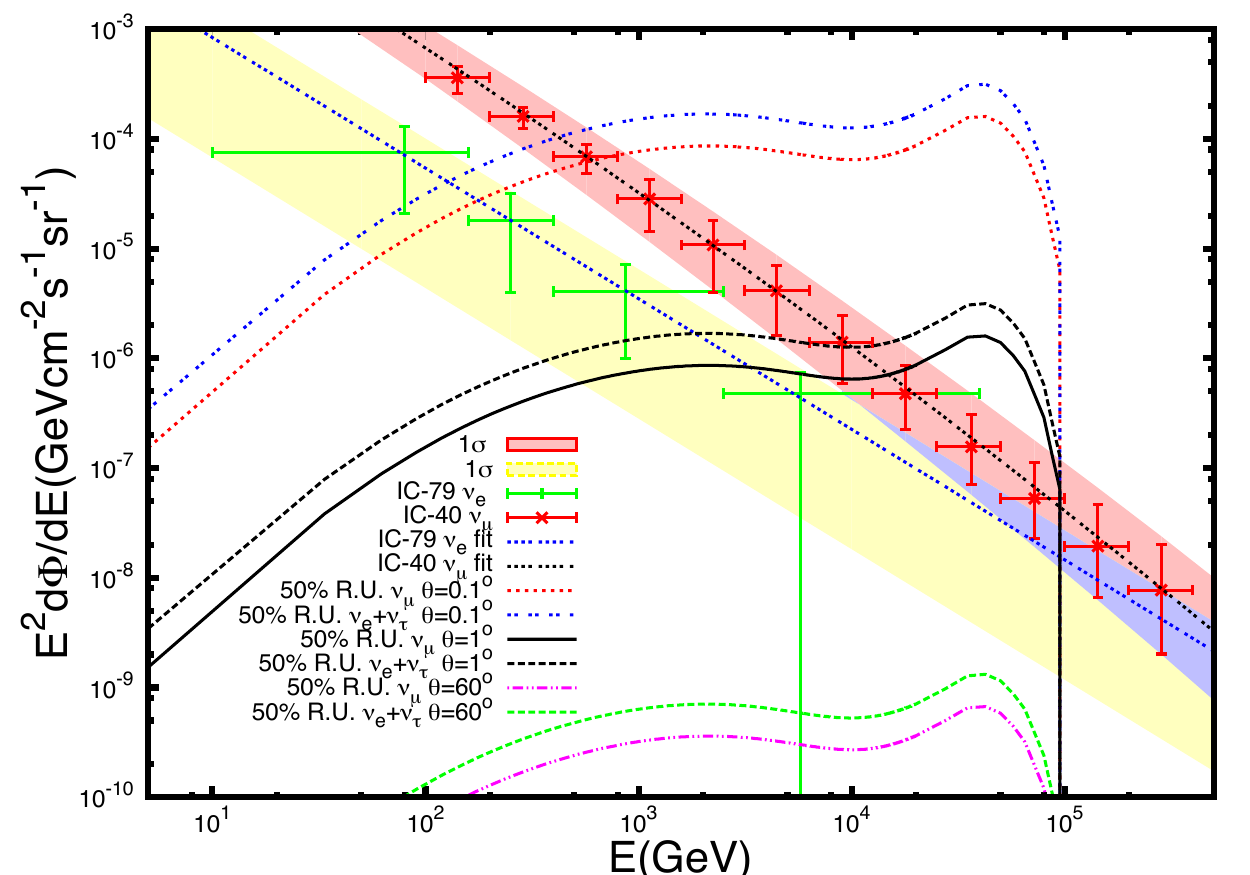}}}
\caption {\footnotesize{
Left panel: The combination of the gamma-ray flux originated by DM $\approx 50$ TeV annihilating into $W^+W^-$ channel and a power-law background well fits HESS and Fermi-LAT data from the GC. An enhancement in the secondary gamma-rays originated by annihilation of DM is needed. An estimation of such enhancement is given throughout the boost factor $b\equiv \langle J_{(2)} \rangle/\langle J^{\text{NFW}}_{(2)} \rangle$ with $\langle J^{\text{NFW}}_{(2)} \rangle\simeq 280 \cdot 10^{23}\; \text{GeV}^2 \text{cm}^{-5}$. 
 The background spectral index $\Gamma\simeq2.63\pm0.02$ is compatible with Fermi-LAT observations \cite{HESS}. Right panel: Expected neutrino fluxes corresponding to muon neutrinos and electron plus tau neutrinos
from $\approx 50$ TeV DM annihilating into $W^+W^-$ bosons for an angular field of view of $\theta=60^\circ$, $1^\circ$ and $0.1^\circ$.
The observed atmospheric muon by the IceCube telescope in the 40-string configuration (IC-40) and electron neutrinos by the 79-string configuration (IC-79)
are also shown together with
the corresponding shared regions at $1\sigma$ confidence level \cite{neutrinos}.
}}
\label{FigWFER}
\end{center}
\end{figure}

\textbf{Gamma-rays.} The differential secondary gamma-ray flux emitted by DM annihilating $100\%$ in $W^+W^-$ SM channel can be easily written as the Eq. (\ref{phigen}) with $\eta_\gamma=1$,  and $\kappa_\gamma=\langle J \rangle_{\Delta\Omega}$. Such gamma-ray flux from DM combined with a power-law background is able to fit HESS data from the GC and Fermi-LAT data \cite{HESS}:

\begin{equation}
\frac{d\Phi_{\gamma-Tot}}{dE}=B^2\cdot \left(\frac{E}{\text{GeV}}\right)^{-\Gamma}+A_W^2 \cdot \frac{dN^{(\gamma)}_{W}}{dE}
\;.
\label{gen}
\end{equation}

where $A_W=\frac{\zeta^{(2)}_W}{2} \cdot \frac{{\Delta\Omega^{\text{HESS}}\,\langle J_{(2)} \rangle}_{\Delta\Omega}}{4 \pi M^2}$. 
The cosmological constraints for thermal DM annihilation cross section is given by CMB observation by PLANCK and WMAP satellite experiments among other measurements  \cite{PLANCK, WMAP}. In the model independent analysis developed here, we assume such value $ \sum^{\text{ SM }}_i \zeta^{(2)}_i \equiv \zeta^{(2)}_W \equiv \langle\sigma v\rangle \simeq 3\times10^{-26}\text{cm}^3\text{s}^{-1}$ and $\Delta\Omega^{\text{HESS}}\approx10^{-5}$. The left panel of Fig. \ref{FigWFER} shows the model independent fit of HESS + Fermi-LAT gamma-ray data.\\

%
%
%


%
%


\textbf{Neutrinos.} The differential flux of neutrinos of flavor $\nu_f$ observed on the Earth 
 can be computed as Eq. (\ref{phigen}) with $\eta_{\nu}\equiv \sum_{p=1}^{3} P_{fp}$. This parameter depends on the dimensions of the source and its distance from the observer: 

\begin{equation}
\frac{d \Phi_{\nu_f}}{dE}=\sum^3_{p=1}P_{fp} \cdot \frac{\zeta^{(2)}_W}{2}
\frac{dN^{(\nu_p)}_{W}}{dE}
 \cdot \frac{{\Delta\Omega\,\langle J_{(2)} \rangle}_{\Delta\Omega}}{4 \pi M^2}
\end{equation}

Neutrinos allow \textit{directional} observation of the GC, although the angular resolution is worse than that for gamma rays. Moreover, it depends on the kind of signal (track or shower) and the position of the source with respect to the horizon of the telescope. All this affects the effective area of the telescope and the possibility to detect a neutrino signal above the background at a given confidence level \cite{neutrinos}. The right panel of  Fig. \ref{FigWFER} shows the detectability of the neutrino signal from $\approx50$ TeV at the GC up the background component, depending on the resolution angle of the telescope. More details on the detectability with $5,\;3$ and $2\sigma $ confidence level may be found in \cite{neutrinos}.\\



\textbf{Antiprotons.} As a difference with \textit{neutral} cosmic-rays that preserve information about the emission direction and spectra, \textit{charged} cosmic-rays propagation in the Galaxy is a complex process affected by many different physical phenomena:
energy losses, convection and reacceleration affect the flux at the Top of Atmosphere (TOA). 
Due to these effects, the direct observation of antiprotons produced at the GC is unlikely. The available antiproton data at TOA
 are in agreement with secondary antiprotons diffusion in the Galaxy \cite{indirect}. In any case, antiprotons generated at the GC might affect the antiproton flux at the TOA. The prospective flux is given by Eq. (\ref{phigen}) with $\eta_{\bar p}=v_{\bar p}/4\pi$ and $\kappa^{(a)}_{\bar p}=\rho^{(a)}R(r_\odot, E_{ \bar p})/M^{(a)}$. For DM totally annihilating into $W^+W^-$ channel the antiproton differential flux at the Solar system is:

\begin{equation}
\frac{\text{d}\Phi_{({\bar p},\odot})}{\text{d}E_{({\bar p},\odot})}=\frac{v_{\bar p}}{4\pi}\cdot\frac{\zeta^{(2)}_W}{2}\frac{\text{d}N^{(\bar p)}_W}{\text{d} E_{\bar p}}\cdot\left(\frac{\rho_\odot}{M}\right)^2R(r_\odot, E_{({\bar p},\odot)})
\end{equation}
The diffusion parameter accounts for two different contribution from the Navarro-Frank-White DM halo and the $\delta$-like DM distribution at the GC:  
\begin{equation}
R(r_\odot,\Ep)=b^W_\text{NFW}\cdot R^\text{NFW}(E_{\bar p})+b^W_{\delta-\text{NFW}}\cdot R^{\delta-\text{NFW}}(E_{\bar p})
\end{equation}
The equivalent boost factor normalization constant is given by
$\langle J\rangle^{NFW}_{\Delta\Omega}\Delta\Omega_\text{HESS}\left(\frac{D_\odot}{\rho_\odot}\right)^2
\simeq 2.13\cdot10^{60}\,m^3\,\text{sr}$ \cite{antipro}. 
We include in the analysis the solar modulation effect on the antiprotons propagation in the Solar system \cite{Perko}. It can be neglected for antiprotons production from heavy DM, because the solar modulation affects energies between $1-10$ GeV \cite{antipro}. More important is the effect of Galactic wind, 
that is described as a convective velocity $V_c$. 
Such velocity at the Galactic disk is $V_c\simeq 0$, 
while at the GC there is evidence of a strong convective velocity of $V_c\simeq 100-1000 km/s$. 
Such effect 
reduces the antiproton flux at low energy. In any case, a detailed model of antiproton diffusion at the GC has not been developed so far. 

\begin{figure}[t]
\begin{center}
\epsfxsize=5cm
\resizebox{6.8cm}{5.0cm}
{\includegraphics{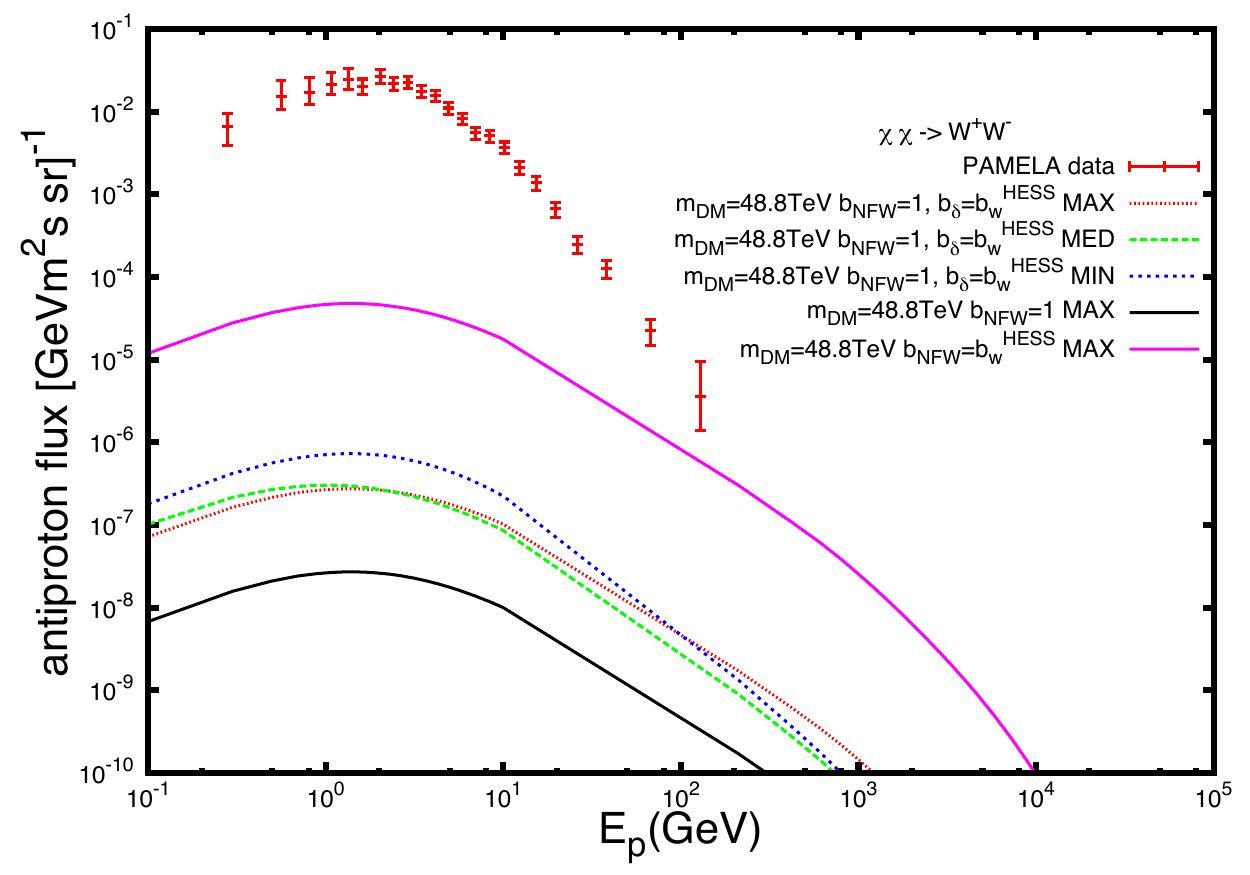}}
\caption {\footnotesize{
Antiproton differential flux at the TOA after propagation for
$\approx 50$ TeV DM annihilating into $W^+W^-$ pairs for different diffusion models and distribution profiles:
The lower signal (black solid line) corresponds to the non-boosted NFW profile by employing the maximum diffusion model.
On the other hand, the same assumptions give raise to the highest flux at high energies (violet solid line)
but with a boost factor of
$b^{WW}_\text{NFW}=b^{WW}_\text{HESS}= 1767$. 
%
We show the antiproton flux at TOA for the medium diffusion model
%
for non-boosted NFW profile 
plus enhanced $\delta$-DM distribution
($b^{WW}_{\delta-\text{NFW}}=  b_{WW}^\text{{HESS}}$)
at the GC for three diffusion models (blue big-dotted, green rushed and red little-dotted line) togehter with PAMELA data \cite{antipro}.
}}
\label{NFWdeltaf}
\end{center}
\end{figure}

\section{Conclusion}
\label{sec-3}
I have reviewed the main results for the study of heavy DM sources at the GC through the observation of different cosmic-rays. The gamma-rays analysis show that HESS and Fermi-LAT data are well fitted with a $\approx 50$ TeV DM candidate annihilating into $W^+W^-$, among other channels. The expected neutrino flux from the same source may be detected after an improvement of the resolution angle of current generation of neutrino detectors. Antiprotons observation at higher energies may also set more stringent constraints on the model. Heavy DM candidate with mass of tens of TeV scale is unconstrained so far. Future investigations will be developed in the direction of both indirect and direct searches, DM model building and deeper understanding of the astrophysics at the GC.

\section{Acknowledgment}
This work was supported by UCM FPI grants G/640/400/8000 (2011 Program), the Spanish MINECO projects numbers FIS2011-23000, FPA2011-27853-C02-01 and MULTIDARK CDS2009-00064 (Consolider-Ingenio 2010 Programme). The author thanks Jos\'e A. R. Cembranos and Antonio L. Maroto for their supervision and participation in the research reviewed in this contribution. 
%

\begin{thebibliography}{}
%
%

\bibitem{indirect}
  A.V. Belikov, G. Zaharijas, J. Silk,  Phys. Rev. D 86, 083516 (2012).
  A. A. Abdo et al. [arXiv:astro-ph.CO/1001.4531v1] (2010).
  M. Chernyakova {\em et.~al.}, ApJ {\bf 726}, 60 (2011).
  T.~Linden, E.~Lovegrove and S.~Profumo,  arXiv:1203.3539 [astro-ph.HE].
  K. Kosak, H. M. Badran, I. H. Bond et al., ApJ, 608, L97 (2004).
  F. Aharonian, A. G. Akhperjanian, K.M. Aye et al. A\&A, 425, L13 (2004b).
  F. Aharonian, A. G. Akhperjanian, K.M. Aye et al. A\&A, 503, 817 (2009).
  J. Albert, E. Aliu, H. Anderhub et al., ApJ, 638, L101 (2006).
  MAGIC collaboration, [arXiv:astro-ph/1103.0477v1] (2011). Maier G. [arXiv:astro-ph.IM/0907.5118v1] . The CTA consortium [arXiv:astro-ph.IM/1003.3703v2], http://www.cta-observatory.org/.



\bibitem{MC}
J. A. R. Cembranos, A. de la Cruz-Dombriz, V. Gammaldi, R. A. Lineros, A. L. Maroto, JHEP 1309 (2013) 077.


\bibitem{antipro}
J. A. R. Cembranos, V. Gammaldi, A. L. Maroto, [arXiv:1410.6689].


\bibitem{Perko}
J.S. Perko, A\&A 184, 119 (1987).


\bibitem{HESS}
J. A. R. Cembranos, V. Gammaldi, A. L. Maroto, JCAP 1304 (2013) 051; Phys. Rev. D 86, 103506 (2012).

\bibitem{branons}
 A.~Dobado and A.~L.~Maroto,  Nucl.\ Phys.\ B {\bf 592}, 203 (2001); 
J.~A.~R.~Cembranos, A.~Dobado and A.~L.~Maroto,  Phys.\ Rev.\ Lett.\  {\bf 90}, 241301 (2003); 
Phys.\ Rev.\ D {\bf 68}, 103505 (2003); 
A.~L.~Maroto, Phys.\ Rev.\ D {\bf 69}, 043509 (2004); 
Phys.\ Rev.\ D {\bf 69}, 101304 (2004); 
Int. J. Mod. Phys. {\bf D13}, 2275 (2004). 
J.~A.~R.~Cembranos  {\it et al.},  JCAP {\bf 0810}, 039 (2008). 


\bibitem{branonsgamma}
  J.~A.~R.~Cembranos, A.~de la Cruz-Dombriz, V.~Gammaldi and A.~L.~Maroto,
  Phys.\ Rev.\ D {\bf 85}, 043505 (2012). 

\bibitem{WMAP}
  E.~Komatsu {\it et al.} [WMAP Collaboration],  ApJ. Suppl. 192, 18 (2011).
  
\bibitem{PLANCK}
Planck Collaboration (Ade, P.A.R. et al.) A\&A (2014) arXiv:1303.5076 [astro-ph.CO].

\bibitem{neutrinos}
J. A. R. Cembranos, V. Gammaldi, A. L. Maroto, Phys. Rev. D 90, 043004 (2014).



\end{thebibliography}
%
%

\end{document}